# The AMMA mulid network for aerosol characterization in West Africa

OLGA CAVALIERI*† , GUIDO DI DONFRANCESCO‡, FRANCESCO CAIRO†,
FEDERICO FIERLI†, MARCEL SNELS†, MAURIZIO VITERBINI †,
FRANCESCO CARDILLO†, BERNADETTE CHATENET§, PAOLA  FORMENTI§,
BEATRICE MARTICORENA§ and JEAN LOUIS  RAJOT ¶

† Consiglio Nazionale delle Ricerche -Istituto di Scienze dell'Atmosfera e del Clima, Italy
‡ Ente per le Nuove Tecnologie Energia e Ambiente, ACS, Italy
§ Laboratoire Interuniversitaire des Systemes Atmospheriques, France
¶ Institute des Researches pour le Developpement, Niger

Three ground based portable low power consumption microlidars (MULID) have been built and deployed at three remote sites in Banizoumbou (Niger), Cinzana (Mali) and M'Bour (Senegal) in the framework of the African Monsoon Multidisciplinary Analyses (AMMA) project for the characterization of  aerosols optical properties. A description of the instrument and a discussion of the data inversion method , including a careful analysis of measurement uncertainties (systematic and statistical errors) are presented.
Some case studies of typical lidar profiles observed over the Banizoumbou site during 2006 are shown and discussed with respect to the AERONET 7-day back-trajectories and the biomass burning emissions from the Combustion Emission database for the AMMA campaign.

## 1. Introduction

Atmospheric aerosols play an important role in the Earth radiative budget via the direct effect by scattering and absorbing solar and terrestrial radiation and via the indirect effect by acting as condensation nuclei of clouds, the principal modulators of solar radiation. Thus, aerosols influence climate, by modifying optical and physical properties of clouds and of the atmosphere as a whole, inducing important changes in their radiative properties, lifetime and precipitation.

*Corresponding author. Email: o.cavalieri@isac.cnr.it

Passive remote sensing instruments on satellites provide a quasi-continuous global picture of the aerosol optical thickness distribution, so that source regions and horizontal transport of the particles may be identified (Kaufman *et al.* 2002). Unfortunately, satellite imagery or passive ground sensors cannot retrieve the vertical distribution of the aerosol layer, which is of importance for aerosol evolution, and which needs to be known for understanding their influence on radiative forcing ( Haywood and Shine 1997) and their effect in clouds formation (Sherwood 2002).

The vertical distribution also affects the residence time in the atmosphere both in the boundary layer and in the free atmosphere, and hence the long range transport of aerosols (Muller *et al.* 2005).

The objective of the African Monsoon Multidisciplinary Analysis (AMMA) project is to improve the knowledge and the understanding of the Western African Monsoon (WAM), its variability on daily to interannual timescales and its influence on the physical, chemical and biological environment regionally and globally. AMMA is a multi-years project involving three observation periods: the long term observation period (LOP) concerned with historical observations to study the interannual to decadal variability and additional long term observations (2001-2010), the enhanced observation period (EOP) planned to serve as a link between LOP and SOP, and the special observation period (SOP).

In particular, the EOP main objective is to document over a climatic transect the annual cycle of the surface and atmospheric conditions and to study the surface memory effects at the seasonal scale. The EOP has a duration of three years (2005-2007).

The SOP periods took place in 2006 and focused on detailed observation of specific processes and weather systems during the dry season SOP0 (January-February) and at various key stages of the rainy season during three periods in summer 2006: the monsoon onset SOP1 (15 May-30 June), the peak monsoon SOP2 (1July-14 August) and finally the late monsoon SOP3 (15 August-15 September).

The identification and characterization of aerosol sources and the study of the evolution and the effects of atmospheric aerosols are among the main objectives of the AMMA effort.

Actually, Africa is the world's largest source of aerosol produced by biomass burning and mineral dust, which in the Sahel constitute the majority of aerosol present in the region (Prospero *et al.* 2002). Satellite imageries show huge plumes of dust and smoke emerging from Africa and spanning the entire Tropical Atlantic and Mediterranean region during much of the year (seawifs.gsfc.nasa.gov/SEAWIFS.html, http://modis.gsfc.nasa.gov/). Mineral dust particles in Saharan desert are lifted up by local convection and transported southwards while biomass burning aerosols from forest fires in the sub-Sahelian region are transported north-eastward, mainly during the dry season in December, January and February ( Marenco *et al.* 1990, Giglio *et al.* 2003, Ito and Akimoto 2007).

Several campaigns have already been carried out in Africa to study aerosols and their optical properties.

Airborne lidar measurements were used to classify the physical and optical properties of Saharan dust observed during the Saharan Dust Experiment (SHADE) (Tanrè *et al.* 2003).

Biomass burning aerosols have been characterized using a variety of ground based and airborne instruments for measuring physical, chemical and radiative properties of aerosols in a field campaign conducted in southern Africa during the Southern African Regional Science Initiative (SAFARI). Schmid *et al.* (2003) reported aerosol optical depth spectra and extinction vertical profiles measured using the NASA AMES airborne 14-channel tracking sun photometer and derived vertical profiles of spectral aerosol extinction.

Magi *et al.* (2003) studied the vertical variability of aerosol scattering and absorption coefficients and Haywood *et al.* (2003) estimated optical properties of biomass burning aerosol by mean of an airborne nephelometer and a particle soot absorption photometer (PSAP).

The AMMA project is active until 2010 and includes measurements at many locations of West Africa and in the Gulf of Guinea, an intensive use of satellite and ground sensor data and several modelling studies. In the framework of this effort, three innovative lidar systems have been deployed along a latitudinal transect over the Sahel to document the transport and evolution of the aerosol from their source regions toward the west of Africa.

Lidar represents a complementary technique to provide vertical profiles of aerosol optical properties with a height resolution of a few meters Gobbi *et al.* 2000, Di Sarra *et al.* 2001, Di Donfrancesco *et al.* 2006 ). A lidar system consists basically of a pulsed laser source firing into the atmosphere, receiving optics to collect the backscattered signal, and detectors and electronics to store the time-resolved atmospheric echo.

The aim of this paper is to describe the instrumental layout and data algorithms for a network of small, portable lidars (microlidars) deployed in Africa within the framework of the AMMA project (Redelsperger 2006) for the study and the characterization of the aerosol properties and evolution over the Sahelian region.

The paper is organized as follows: section 2 describes the instrument hardware and performances, section 3 is devoted to a discussion on the data inversion algorithm, section 4 deals with the error analysis, section 5 explains field instruments deployment and ancillary measurements while in section 6 some interesting case studies from the database are presented. Finally in section 7 some conclusions from our work are presented.

**2. MULID systems**

The micro lidar (MULID) systems are newly developed portable low power consumption lidar systems adapted for a field use in a semiautomatic mode in remote sites powered only by solar panels

The MULID lidar system employs a small, low-weight flash pumped Nd: YAG laser (Kigre Inc.) as laser source, with second-harmonic generation and passive Q switching. The laser pulse duration is 10 ns , with an output energy of 5 mJ/pulse at 532 nm. The maximum repetition rate of the passively Q-switched laser is approximately 0.3 Hz, but, to extend the laser lifetime, it has been reduced to 0.1 Hz.

The four times expansion of the laser beam with a small expander, reduces the full angle divergence of the beam to about 0.5 mrad. This high-power, low repetition rate laser has several advantages such as compact size of the laser head (15× 2×2 cm), low weight (0.35 Kg) and ultra low power consumption (< 1 W). The lidar is to be considered eye safe after a distance of 1.5 Km. The choice of a high energy – low pulse repetition rate laser was worthwhile, since it assures a high signal to noise ratio, with respect to the "high pulse rate - low pulse energy" laser approach used in most compact lidar systems. Our choice renders daytime measurements easy even with large field-of-views (FOV) telescope, which is mandatory considered by the need to obtain lidar measurements from low altitudes and the requirement to have an easy and stable alignment. The atmospheric echo is collected by a Newtonian 20 cm diameter f/1.5 telescope, subsequently collimated and spectrally filtered by a narrowband interference filter with 2 nm bandwidth (Semrock model). These filters, although not extremely narrow, have a high signal transmission ( > 90%) and an extremely low temperature dependence which make them particularly adapted to field measurements where high temperature gradients are expected.

The telescope FOV was set to 0.67 mrad by using a field stop of 200 micron, with the laser beam - telescope FOV overlap starting at around 40 m from the instrument and a complete overlap above 600 m , where also the defocusing becomes negligible.

A polarizing beam-splitter cube is used to split the optical signal in two components, one polarized parallel to the laser light, the other polarized perpendicularly. Each component is detected by a miniaturized photomultiplier module (Hamamatsu Model 5783P) with very low thermal noise (below 10 counts/s at 25°C). The signal of the atmospheric echo from the photomultipliers is acquired simultaneously in analog and photon-counting modes. The acquisition electronics (Embedded Devices, APC-80250DSP) is based on a FPGA technology and uses a fast digital signal processor (DSP) for analog as well as photo counting signal averaging and processing. Online averaging with an integration time which can be set via software ( in our case set at 10 minutes) to give a single profile, is provided. Successively the averaged profile is stored in the onboard memory. The signal acquired in analog mode via the A/D converter has a spatial resolution of 3.75 m from ground level up to 3.75 Km. The photon-counting profile has a coarser resolution of 30 m from ground level up to 30 Km of altitude.

The use of photon-counting detection is imperative in the acquisition of echoes from the far range, while a fast A/D converter is necessary for the echo from the first kilometres since the atmospheric echo always remains well in the analog detection regime. By averaging few tens of subsequent atmospheric echoes (i.e. 10 minutes), a profile up to 6 Km is easily obtained in daylight conditions.

In our measurement conditions, the photo counting data were exploitable above 3 Km, since below that level, especially during daylight, photon counting digital signal often saturates. Nevertheless the region of overlap between the analog and photo counting signal is large enough to allow a reconstruction of the signal profile from 600 m up to 30 Km. After the subtraction of the background noise, a common profile made from the analogical signal in the lower part, and the digital signal in the higher part, is created. The region where the two signals can be superimposed generally locates between 3 Km and the top of the analog profile at 3.75 km, a region where the photo counting detection has a good linearity and the analog detection is still sufficiently sensitive.

The system installed in Banizoumbou has also an additional channel at 1064 with an avalanche photodiode (APD) as detector . The long wavelength enhanced silicon APD sensor (C30954-EG&G) is operated in current mode. An electronic circuit has been developed (Embedded Devices) to keep a constant gain over a large temperature range by adjusting the polarization voltage of the sensor.

Moreover , a close range 532 nm total elastic channel characterized by a lens diameter of 25 mm , a field of view of 2 mrad and f =25 mm is also available in the Niger system. The low altitude channel is used to extend the lidar profile down to 30 m.

The possibility to reconstruct the lidar profile from about 20 meters above the ground up to the top of the atmosphere assumes great interest when a comparison between the lidar data and various complementary aerosol measurements taken at the ground is performed. Moreover, the use of a second wavelength is valuable for the discrimination of various types of aerosols, since it has been demonstrated that colour index provides information on the average particle size (Rajot *et al*. 1994).

The systems were hosted in huts with windows facing the sky, and experiences large temperature gradients in the diurnal cycle.

The MULIDs were thermally insulated from the hot environment by using a polyurethane shield. An additional Peltier cooler (30 W) with fans was installed to keep the internal temperature 10-15°C below the environmental temperature resulting in operational temperature between 15 and 35 °C. The overall power consumption is below 100W, so the systems can be easily powered by solar panels and batteries. The systems automatically perform two measurement sessions per day, at fixed hours, one in the morning and one in the evening, each session lasting one hour. During SOPs, additional measurement sessions, activated by an operator present on the site, were performed often in coincidence with other campaign activities, such as aircraft over passes. The data are stored in onboard mass memory which have to be downloaded offline by an operator every week.

The polarization sensitive observations of the MULID systems allow reconstructing aerosol backscattering and depolarization profiles useful to characterize the optical properties of the aerosol and to discriminate layers of particles of different characteristics that may indicate differences in their origins and evolution.

In Table 1 the main optical and electronic MULID characteristics are listed.

**3. Data inversion algorithm**

In a Rayleigh polarization lidar system, a polarized laser beam is emitted, backscattered in the atmosphere and collected by two detection channels with selection of parallel and cross-polarized backscattering.

The powers measured on both channels from range z, $P^{//}$ and $P^{\perp}$, are linked to the quantity of physical interest, the volume backscatter coefficient $\beta$ of the sampled air at altitude $z$ by the single scattering LIDAR equation expressed as :

$$P^{//,\perp}(z) = P_0 (c\tau/2) C^{//,\perp} \frac{\beta^{//}(z) + \beta^{\perp}(z)}{z^2} \exp[-2\int_0^z \alpha(\xi)d\xi] + P^{//,\perp}_b \tag{1}$$

where $P_0$ is the transmitted pulse energy, $c$ is the speed of light, $C$ is the gain factor of the detection system. The symbols $//$ and $\perp$ indicate parallel and perpendicular polarized components (with respect to the laser polarization plane), respectively, of the lidar signal. $\beta$ is the volume backscattering coefficient, $\alpha$ is the extinction coefficient and $\tau$ is the laser pulse duration. $P^{//,\perp}_b$ is the background from sky light on the parallel and cross polarized receiving channel respectively.

The speed of light links the time and space resolved measurements, being $z=ct/2$, where $t$ is the time elapsed since the laser firing.

An important optical parameter can be evaluated directly from the powers recorded on the two channels, the volume depolarization ratio, defined as the ratio between the backscattered signal component with polarization perpendicular and parallel to the laser polarization plane, or, in terms of volume backscatter coefficients, as:

$$D = \frac{\beta^{\perp}}{\beta^{//}}$$

This expression is directly linked to the observed powers collected in the parallel and cross polarized channel by:

$$D(z) = K * (P^{\perp} - P^{\perp}_b)/(P^{//} - P^{//}_b)$$

The coefficient $K$, a calibration constant accounting for the difference in receiver responses for the two channels, is chosen in order to obtain the theoretical value to be expected from the atmosphere in a region where the aerosol contribution to the backscattering can be considered negligible (Young 1980). In our case, this theoretical value was set to 0.014 (Behrendt and Nakamura 2002).

The volume depolarization ratio, apart from the fixed known contribution from the air molecules, has a variable contribution from the suspended aerosols and clouds, the latter being a function of the number, shape and size of the particles.

Since only non-spherical particles can produce a change in the polarization plane of the backscattered light, the volume depolarization ratio *D* provide an indication of the particle shape allowing to distinguish between spherical, small non spherical (i.e. biomass burning aerosol) and large non-spherical particles (i.e. mineral dust). Values close to zero are expected for spherical particles, and higher values are produced by non spherical particles (Reagan *et al.* 1989). Moreover, up to a certain extent, higher values of depolarization are signature of larger particles (Mishenko *et al.*1997)

It is important to stress here the straightforward computation of the volume depolarization which is simply the ratio of powers detected from the two receiving channels, assuming that the extinction due to molecules and aerosols equally affect the parallel and cross backscattering.

An important source of systematic error in depolarization assessment comes from the splitting in the detection of the parallel and the cross polarizations, which leads to a mixing in the signal or the cross talk between receiving channels (Cairo *et al.*, 1999).

The cross talk error has been taken into account and the volume depolarization profiles have been corrected for a cross talk of 2.5 % between channels, as reported and measured on the polarizing beam splitter cube.

As a control, we verify the volume depolarization value in the regions of lidar profile where ice clouds are present and we found the typical value as found in literature ( Snels *et al.* 2008).

Since the laser light is scattered by air molecules and particulate matter, the volume backscattering coefficient $\beta$ can be split into contribution from molecules ( $\beta_m$) and aerosols ($\beta_a$):

$$\beta(z) = \beta_m(z) + \beta_a(z)$$

and so the extinction coefficient $\alpha$:

$$\alpha(z) = \alpha_m(z) + \alpha_a(z)$$

where, from Rayleigh molecular scattering theory, the molecular extinction coefficient $\alpha_m(z)$ equals $\frac{8}{3}\Pi \beta_m$.

The molecular backscatter $\beta_m$ and hence the molecular extinction coefficient $\alpha_m$ can be evaluated by the Rayleigh scattering theory from air density obtained from radio sounding or from a suitable atmospheric model, while the aerosol contribution to backscatter and extinction coefficients is unknown, and these quantities are to be retrieved from the measurement.

The total profile $P(z)$ is calculated as $P(z)=P^{//}(z)(1+D)$, where $D$ is the depolarization ratio.

A common quantity used in literature to express the atmospheric aerosol burden is the backscattering ratio $R$, defined as the ratio of the total volume backscattering coefficient (molecular plus aerosol) to the molecular backscattering coefficient:

$$R=(\beta_m+\beta_a)/\beta_m$$

Since both the aerosol backscatter and extinction coefficient are present in the lidar equation (1) as variables, we have to assume some kind of relation between $\alpha_a$ and $\beta_a$, the so-called extinction to backscattering ratio, or "lidar ratio", in order to invert the lidar equation:

$$L=\alpha_a/\beta_a$$

Let $S(z)=P(z)*z^2$ be the range corrected and background subtracted signal, and let the total attenuated backscattering ratio $R'$ be defined as $R'(z)=K_R*S(z)/\beta_m$, where the molecular backscattering $\beta_m$ has been derived from radio sounding density profiles by means of the formula (Hinkley, 1976):

$$\beta_m = N_m \frac{d\sigma_m(\pi)}{d\Omega}$$

where $N_m$ is the number of molecules per unit volume ($N_m = \frac{P}{K_B T}$ where $P$ is the atmospheric pressure, $T$ is the atmospheric temperature and $K_B$ is the Boltzmann constant) and $\frac{d\sigma_m(\pi)}{d\Omega}$ is the differential Rayleigh scattering cross section.

Since $R'(z)= 1$ in absence of aerosols, the coefficient $K_R$ has been suitably chosen to impose to the attenuated backscattering ratio a value equal to one in a region of the atmosphere supposedly free of aerosols.

In order to accomplish this task, the molecular profile is adopted as a reference profile for the aerosol free portion of the corrected and background subtracted signal.

The total backscattering ratio is obtained by iteratively correcting $R' =R'_i$ for cloud, aerosol and molecular attenuation until the process converges to a stable value R. In order to do that, a lidar ratio $L$ value has to be assessed. As we have stated above, the correction for the molecular attenuation is straightforward and descends from Rayleigh theory, while a different approach is needed for aerosol and cloud particles. Our choice was to fix the lidar ratio to constant but different values both in regions of the atmosphere where clouds were present and where aerosol are present.

We classified as liquid and ice clouds regions of the profile where particular conditions for the backscatter ratio, depolarization ratio and altitude are met and for these region we fix the respectively lidar ratio values (Chen *et al*. 2002, O'Connor *et al*. 2004). The regions in which large aerosol events, or mixed phase clouds, are occurring are also considered. However this latter event resulted to be rare compared to the presence of aerosols, and we limited ourselves to the consideration of then sole aerosol occurrence. So, in general we considered these cases as aerosols, and we used a suitably chosen constant value for the lidar ratio along the vertical profile. This classification is summarized in table2.

Thus, among the various aerosol types more likely to be observed on our sites, we can say that desert dust aerosol is characterized by lidar ratio values $L$ from 30 to 50 sr and volume depolarization greater than 10% while biomass burning aerosol by a lidar ratio value of ~60-70 sr and volume depolarization often lower than 10 % ( Murayama *et al.* 1999, Ferrare *et al*. 2001, Fiebig *et al*. 2002).

Table.3 is shown a list of the typical lidar ratio values obtained for desert dust and biomass burning aerosols found in literature.

We used the Niger lidar dataset, where a full profile was available from 20 m upward, to test different approaches to the lidar inversion procedure, by using the co-located measurements of AOD from the AERONET sunphotometer to better constrain the lidar ratio.

In this work, we used AOD data from level 2.0 and we interpolated the data at 440 and 675 nm to obtain the equivalent AOD at 532 nm, the wavelength of lidar measurements.

The lidar AOD was achieved by integrating the extinction profiles measured by MULID up to 6 Km of altitude and compared to the aerosol optical depth (AOD) measured by the

AERONET sun photometer at Banizoumbou station.

We used for this comparison only the MULID evening profiles because in the morning the signal from the low altitude channel saturates.

Moreover, due to the fact that a limited measurement period of MULID has to be related to the continuous AERONET measurement series, the AOD data closer in time to the MULID data are chosen for the comparison.

We tested a depolarization-dependent lidar ratio formulated in order to meet the values expected in the cases of biomass burning and desert dust aerosols, classified according to the values of $L$ reported in table 3. This approach was compared with inversions carried out with lidar ratios kept constant along the profile, with their values ranging from 30 to 70 sr. For Cinzana and M'Bour, a fix value of the lidar ratio equal to 40 sr resulted in a better accordance of the two AOD - on the average - compared with the one resulting from the use of the depolarization-dependent lidar ratio.

For Banizoumbou, we allowed the inversion procedure to seek for the value of the lidar ratio that accomplished the best agreement with the co-located AOD measurements case by case. The results of this comparison are shown in figure 1.

The AOD values obtained by Calipso observations and those obtained by integrating the scattering coefficient profile measured by the airborne AVIRAD aerosol sampling system are also shown.

The series have a similar trend but a systematic bias. The lidar AOD and the nephelometer AOD are in good agreement underestimates the AERONET AOD along the entire period. Although the cause of this discrepancy is still unclarified, different causes might be invoked:

- a different observation geometry of the two instruments;
- different vertical range of integration in AOD computing, from 20 m to 6 km in the lidar case, from the ground to the top of the atmosphere for the sunphotometer;
- different spatial position of sunphotometer respect lidar location.

## 4. Error analysis

The assumption of a constant lidar ratio is an important source of systematic errors in the calculation of backscattering ratio as it is documented in the literature ( Sasano *et al.* 1985). Figure 2 shows the backscatter ratio profiles calculated for a particular case study (julian day 148) in the Niger site, for a particularly large aerosol burden in the atmosphere. The displayed profiles have been calculated using different lidar ratio values. The two values at 70 sr and 30 sr for the lidar ratio represent the extremes of the uncertainty to be attributed to the backscattering ratio parameter. The dashed line represents the relative error on the mean profile due to undetermined lidar ratio value. The uncertainty on the whole profile, which - due to the heavy aerosol burden - represent a sort of worst case, maximizes near the ground to a value of 27%, decreasing upward to a value of 10% at 2.5 km and becoming negligible above 3.5 km, where the aerosol profile has been calibrated.

Apart from uncertainties to be attributed to the choice of the lidar ratio, lidar measurements are subject to a number of other experimental uncertainties, arising both from the measurement and calibration errors. The error propagation theory is usually used in the error analyses of lidar observations yielding algebraic expressions for the probable error as a function of various error sources (Liu *et al.* 2006, Russel *et al.* 1979)

We estimated for the same case study (julian day 148) the error for both parallel and cross polarized MULID signal and for the volume depolarization profile D.

The errors are calculated on the analog and digital signals ($\frac{\Delta P_A}{P_A}$ and $\frac{\Delta P_D}{P_D}$) respectively for both polarization as:

$$\frac{\Delta P_A}{P_A} = \frac{2\Delta S}{P_A}$$ where $2\Delta S$ is the semi dispersion on the mean signal and $P_A$ the analog signal background subtracted;

$$\frac{\Delta P_D}{P_D} = \frac{\sqrt{n}}{n} + \frac{\sqrt{n_b}}{n_b}$$ with n the number of counts (background subtracted) and $n_b$ the number of counts of the background ;

For the volume depolarization profile the error is computed as:

$$\frac{\Delta D}{D} = \frac{\Delta P^{\perp}}{P^{\perp}} + \frac{\Delta P^{//}}{P^{//}}$$

The error profiles are shown in figure 3.

The error on the parallel polarized signal (black line) is less than 12% along the entire profile, with minimum values (< 4 %) in the low levels up to 3 Km and the maximum in the layer between 3 and 4 Km.

The error on the cross-polarized lidar signal (purple line) is larger than the parallel polarized one up to about 3 Km of altitude with values < 10 %. In the layer between 3 Km and 4 Km the error values are around 8 %.

The error on the depolarization D (red line) is less than 12% below 2.5 Km of altitude and reaches the maximum value in the layer between 2.5 and 4 Km with values of about 18%.

## 5. Field deployment and ancillary measurements

Three micro lidar systems have been deployed (since January 2005) along a longitudinal transect named Sahelian Dust Transect which represents the main transport pathway from the dust source regions toward west. The MULIDs operated in three stations located in M'Bour (Senegal, 14.23°N-16.57°W), Cinzana (Mali, 13.16°N-5.56°W) and Banizoumbou (Niger, 13.5°N-2.6°E) for SOP and EOP operations.

In Cinzana the lidar is located close to the buildings at about 2 m above the ground while in M'Bour station it is positioned on the roof of a small building at 10 metres from the ground. Both are rural stations, surrounded by open fields. Along with the lidar systems, the sites host a Tapering Element Oscillating Microbalance (TEOM) measuring particle mass concentration at ground level (in µg m$^{-3}$), an automated collector monitoring the wet and dry deposition and a photometer. The M'Bour site, being one of the three super-sites implemented in Africa for the AMMA activities, is also equipped with particle counters, nephelometers, aethalometers, fluximeters and photometers.

The station in Banizoumbou, located in the countryside at a distance of 60 km east from the capital of Niger, Niamey, is a second super site for the AMMA project, hosting instruments for a complete characterization of the aerosol properties. This site has been operational since the early nineties, when the first measurements of soil erosion were performed on a cultivated field and a fallow ( Balis *et al*.2004, Rajot 2001). For the complete list and description of instruments placed in the Banizoumbou and M'Bour super sites the reader may refer to: http://amma.mediasfrance.org/implementation/instruments/.

The field campaign activities are supported by a network of passive radiometers which has been routinely operative in West Africa since 1995, within the activities of the Aerosol Robotic Network (AERONET). This network of annual sky calibrated radiometers measures the direct solar radiance at eight wavelengths and sky radiance at four of these wavelengths, providing sufficient information to determine the aerosol size distribution and refractive index

( Dubovik *et al.* 2000, Dubovik *et al.* 2002). This long-term, continuous and readily accessible database of aerosol optical, microphysical and radiative properties for aerosol research and characterization is available at http://aeronet.gsfc.nasa.gov/. The AERONET (Aerosol Robotic NETwork) program provides aerosol optical depth (AOD) and Angstrom coefficient (the ratio of the AOD at two different wavelengths) data for three data quality levels: level 1.0 (unscreened), level 1.5 (cloud-screened) and level 2.0 (cloud-screened and quality -assured) in several wavelength bands (440 nm, 675 nm, 870 nm and 1020 nm). Almost daily data are available at the Banizoumbou site for the year 2006, from about 7 am to 5 pm.

The Calipso data used in this work are the CAL_LID.L240KmAProcal Beta-V201 data available at http://eosweb.larc.nasa.gov.

The satellite was launched on 28 April 2006 and is now flying as part of the Afternoon Constellation (Atrain) , consisting of five satellites, all flying at an altitude of 705 km in a sunsynchronous polar orbit, with an equator crossing time of about 1:30 PM local solar time.

CALIOP is a three-channel backscatter lidar optimized for aerosol and cloud profiling. Linearly polarized laser pulses are transmitted at 532 nm and 1064 nm. The 1064 nm receiver channel is polarization insensitive, while the two 532 nm channels separately measure the components of the 532 nm backscatter signal polarized parallel and perpendicular to the outgoing beam.

Presently, the dataset time coverage extends from day 180 of 2006 to day 260 of 2008. Profiles correspond to satellite passages at around 1.30 GMT and 13:00 GMT.

For each satellite orbit in the area 11°N-15°N and 0°W-5°W , a single profile of total backscatter coefficient and aerosol extinction coefficient has been created by averaging all Calipso measurements in the area, related and representative of the aerosol content in the Niger site.

In situ observation of vertical profiles of aerosol physico-chemical and optical properties during monsoon season 2006 were performed onboard the SAFIRE ATR 42 (F-ATR42).

The F-ATR42 was based in Niamey and equipped with standard instrumentation for aerosol sampling and characterisation of the aerosol physico-chemical and optical properties. This was done from a newly designed inlet, AVIRAD.

During the SOP1 and SOP2 period, the ATR-42 has performed 23 vertical profiles. The majority of those have been performed above the ground-based station of Banizoumbou in order to coordinate with the lidar and in-situ measurements performed at this supersite.

The vertical aerosol scattering coefficient profiles at 550 nm performed by the nephelometer onboard the F-ATR42 above the ground based station of Banizoumbou were compared with the simultaneous aerosol extinction coefficient profiles acquired by the MULID system.

The identification of the origin of air masses arriving at the three stations and for a characterization of the different kind of aerosol present in the atmosphere and observed by lidar, the AERONET 7-day back trajectories ending respectively on Banizoumbou, Cinzana and M'Bour sites have been calculated at different height levels: 0-2 Km, 2-6 Km and above 6 Km of altitude.

Finally, the biomass burning emissions from the Combustion Emission database for AMMA campaign have been used.

## 6. Desert dust and biomass burning observations

A great number of vertical profiles are available for 2006, in particular during the monsoon onset (May-June) and the peak monsoon (July-August) : 157 profiles from February to August 2006 in the Niger site and 48 profiles from January to July 2006 for the Mali site are available but due to problems with laser lifetime some periods were not covered by measurements.

A review of the aerosol profiles in the Niger site shows as the most prominent feature the presence of desert dust aerosols between ground up to about 4-5 Km of altitude with high value of depolarization (>10%).

Very few profiles (10% of the whole dataset) instead show the presence of two aerosol types : a desert dust layer in the lower levels up to 2- 3 km of altitude and an elevated layer of biomass burning aerosols at altitudes between 3 and 6 Km.

Occasionally mixing of the two aerosol type took place always between 3-6 Km of altitude above a desert dust layer.

Together with the backscatter and extinction profiles, for the identification of the origin of the air masses arriving at the three stations and for a characterization of the different kind of aerosol present in the atmosphere and observed by lidar, the AERONET 7 day back trajectories ending respectively on Banizoumbou, Cinzana and M'Bour sites have been calculated at different heights levels and also the biomass burning emissions from the Combustion Emission database for AMMA campaign have been used.

We can distinguish the lidar profiles where biomass burning might be present from those where on the contrary there is only dust using both the information from back trajectories and from the biomass burning emissions map. When the back trajectories pass over biomass burning sources before arriving on the site where the lidar system is placed we can say that in these lidar profiles biomass burning might be present. On the contrary, when no back trajectories pass over biomass burning sources, we are quite sure that the lidar profiles are characterized by the presence of only dust.

Three cases of typical lidar profiles observed over Banizoumbou during the monsoon season 2006, characterized by different distributions of mineral dust and biomass burning aerosol are shown in figure 4, while in figure 5 are shown the AERONET 7 day back trajectories ending on Banizoumbou station in the three case studies at different height levels and the biomass burning emissions .

On June 06 ( see figure 4(*a*)) in the morning a layer of particles can clearly be identified between ground and 1 km of altitude by its high volume depolarization ratios (30-40 %) and high extinction values greater than 0.4 Km-1.

These values are consistent with the hypothesis that the particles in these layers consist of mineral dust particles. This hypothesis is confirmed by the analysis of back-trajectories, showing how the origin of these air masses are far from the regions where biomass burning took place, while some back-trajectories between ground and 6 km of altitude indicate airflow coming from the Saharan desert in the North, source region for mineral dust ( see figure 5(*a*)).

A case of two layer profile is observed on May 22 (see figure 4(*b*)) in the evening. An aerosol layer of a few hundred thickness meters is observed between ground up to about 0.5 Km of altitude characterized by high extinction values around 0.2 Km-1 and volume depolarization greater than 25% along the entire layer.

Above 2 km of altitude there is another aerosol layer characterized by lower values of volume depolarization around 10% and extinction values around 0.1 Km-1.

Thus, the layer close to the ground can clearly be identified as dust while the particles in the layer between 2 and 6 Km of altitude can be recognized as biomass burning aerosols. This is confirmed by air mass back-trajectory analysis which show how the back-trajectories at each altitude correspond with air masses arriving from the east passed over a region characterized by large emission of biomass burning (see figure 5(*b*))

On May 27 (see figure 4(*c*)) in the evening another two layers profile was observed; the first layer extended from ground up to 0.7 km of altitude with large values of extinction reaching more than 0.2 Km-1 at 0.5 Km and depolarization ratios around 30 %. These values suggest that this layer is dominated by mineral dust particles.

The layer above 2 km is characterized by quite high extinction values around 0.2 Km-1

typical for dust layers but the depolarization ratios is significantly lower (12-13 %) and reaches values observed for biomass burning layers. Thus, these values indicate a mixture of dust and biomass burning aerosol. Indeed the air masses arriving at altitude between ground and 6 km of altitude mainly arriving from the North or stay close to the location passing on some biomass burning emission sources (see figure 5(*c*)).

**7. Summary and Conclusion**

In this paper we have presented the micro-lidar (MULID) network built and deployed in the framework of the AMMA project in three remote sites in Niger, Mali and Senegal, for the study and the characterization of the aerosol properties and evolution over the Sahelian region.

These systems contribute to AMMA providing regular observations of aerosol backscattering and depolarization profiles: during the year 2006 and especially during the Special Observing Period (SOP) a great number of profiles predominantly in Niger and Mali site have been collected and analysed.

A description of the instrument, its capabilities and limitations, is presented and the data inversion method is discussed in detail.

Three case studies of typical lidar profiles over Banizoumbou during year 2006 are presented and discussed using also the AERONET 7 day back trajectories ending on Banizoumbou site and the biomass burning emissions map available on AMMA database.

For the case of Niger, we compared AOD data from AERONET database and the Mulid AOD calculated using different lidar ratio values characteristic of the type of aerosol present in the area. From this comparison it is clear that there is a systematic bias between the two time series of measurements: the trend is similar but lidar data underestimate the AOD from AERONET database along the entire period.

We believe that these small, low-power, portable instruments represents a good tool for aerosol characterization along the Sahelian dust transect and allow possible comparison with other ground-based and aircraft measurements for a complete analysis of the chemical and physical aerosol properties.

**Acknowledgments**
Based on a French initiative, AMMA was built by an international scientific group and is currently funded by a large number of agencies, especially from France, the United Kingdom, the United States, and Africa. It has been the beneficiary of a major financial contribution from the European Community's Sixth Framework Research Programme. Detailed information on scientific coordination and funding is available on the AMMA International Web site at www.amma-international.org.

Table 1. Mulid System Parameters

| Parameter | Value |
|---|---|
| Wavelengths | 1064 nm and 532 nm |
| Laser Type | Nd: YAG |
| Pulse duration | 10 ns |
| Laser repetition rate | 0.3 Hz |
| Laser output energy | 10 mJ at 1064 nm |
|  | 5 mJ at 532 nm |
| Telescope diameter | 20 cm |
| Telescope type | Newtonian f/1.5 |
| Telescope field of view | 0.666 mrad |
| Beam divergence | 0.5 mrad, full angle |
|  | x 4 expanded |
| Effective Filter bandwidth | 2 nm |
| Raw data resolution | 30 m photo-counting |
|  | 3.75 m analog mode |
| Processed data resolution | 30 m |

Table 2. Classification of different type of aerosols depending on backscattering ratio R and depolarization D values and the corresponding lidar ratio value L chosen .

| Type of aerosol | R | D (%) | L (sr) |
|---|---|---|---|
| Ice clouds | R>10 | D>20 | L=30 |
| Water clouds | R>10 | D<2 | L=19 |
| Mixed phase clouds/ large aerosol events | R>10 | 2 < D < 20 | L=40 |

Table.3 Lidar ratio values for desert dust and biomass burning aerosol from literature.

| Aerosol type | λ (nm) | Lidar ratio (sr) | References |
|---|---|---|---|
| Dust | 355 | 37-57 | Mona et al. (2006) |
| Dust | 351 | 50-55 | De Tomasi et al. (2003) |
| Dust | 440 | 50-70 | Tafuro et al. (2006) |
| Dust | 532 | 40-75 | Mattis et al. (2002b) |
| Dust | 532 | 45 | Immler and Schrems (2006) |
| Dust | 550 | 42 ± 4 | Cattrall et al. (2005) |
| Biomass burning | 355 | 62 | Balis et al. (2003) |
| Biomass burning | 532 | 40-80 | Wandinger et al. (2002) |
| Biomass burning | 550 | 60 ± 8 | Cattrall et al. (2005) |

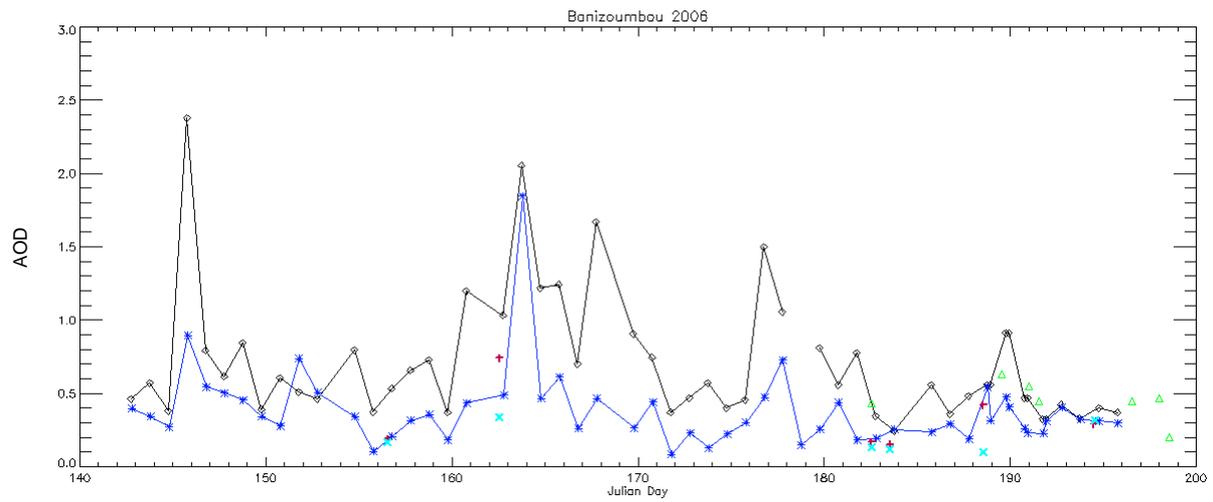

Figure 1.The AOD derived from Mulid extinction profiles (blue line and asterisks *) compared to the AERONET sun-photometer AOD (black line and diamonds ◊) measured at Banizoumbou station during 2006. Are also shown the AOD values from Calipso data (green Δ), the AOD data from aerosol scattering coefficient profiles performed by nephelometer AVIRAD aerosol system onboard the ATR42 above Banizoumbou site around midday (magenta +) and the lidar AOD data simultaneous with airborne measurements (cyan x)

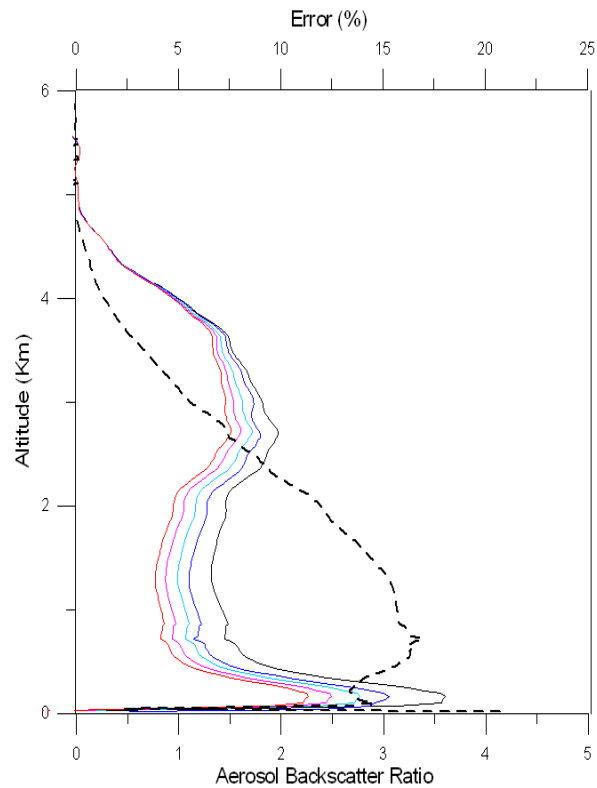

Figure 2. Aerosol backscatter ratio profiles calculated for a particular case study (julian day 148) in the Niger site using different lidar ratio values : L=30 sr ( black line), L=40 sr (blue line), L=50 sr (cyan line), L=60 sr (magenta line), L= 70 sr (red line) . The dashed line represents the relative error computed on the mean profile.

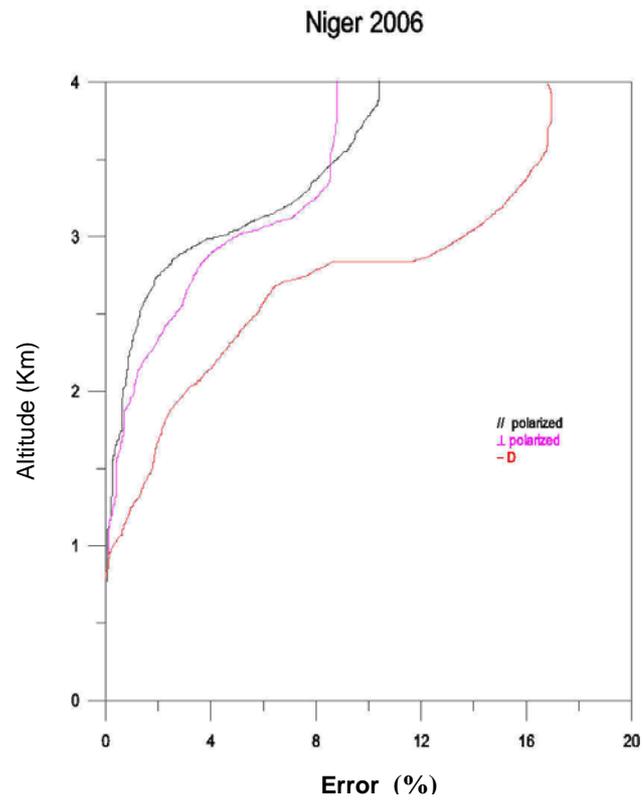

Figure 3. Error profiles calculated using the error propagation theory for a particular case study in the Niger site (Jday=148) on the parallel polarized lidar signal (black line), on the cross-polarized lidar signal (magenta line) and on the volume depolarization profile D (red line).

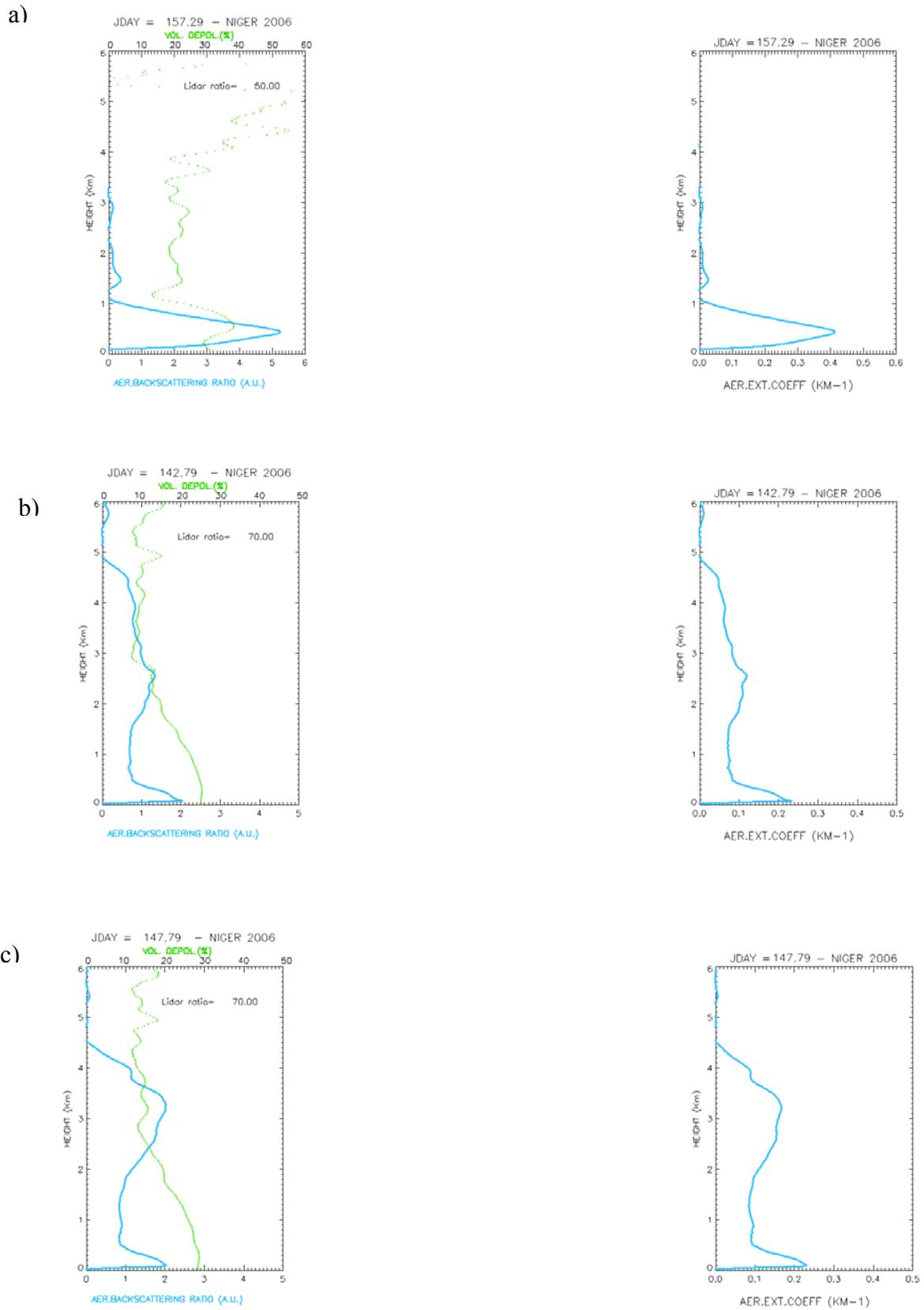

Figure 4. Aerosol backscattering ratio (left) and aerosol extinction coefficient (right) profiles on June 6, 2006 (a), May 22, 2006 (b) and May 27, 2006 (c).

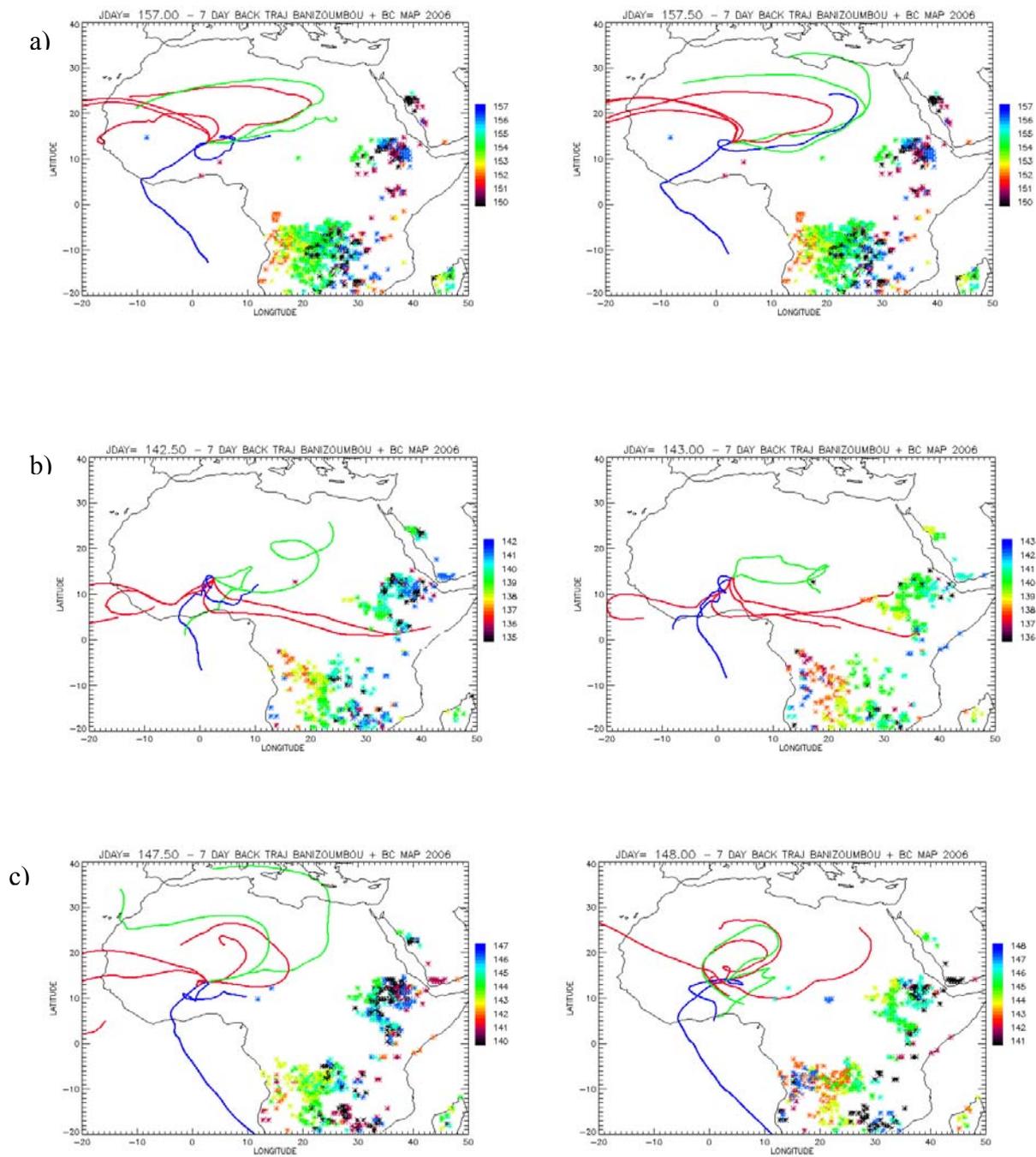

Figure 5. Aeronet 7 days back-trajectories ending at Banizoumbou station on June 6, 2006 (a), May 22, 2006 (b) and May 27, 2006 (c) at different height levels : blue trajectories between ground and 2 Km, green between 2 and 6 Km and red above 6 Km of altitude.

The ∗ symbols represent the biomass burning emissions from the Combustion Emission database for AMMA campaign available on http://www.aero.obs-mip.fr:8001/ and the different colours indicate the day of the emission.